\documentclass{PoS}

\usepackage{mathpazo}
\usepackage[utf8]{inputenc}
\usepackage[english]{babel}
\usepackage{amsmath,amssymb}
\usepackage{mathrsfs}
\usepackage{graphicx}
\usepackage{float}
\usepackage[margin=1cm]{caption}
\usepackage{feynmp}
\usepackage{cite}

\usepackage{tikz}
\usetikzlibrary{arrows.meta}

\title{Chiral Perturbation Theory with an Isosinglet Scalar}
\ShortTitle{Chiral Perturbation Theory with an Isosinglet Scalar}

\author{
\speaker{Martin Hansen} \\
INFN Roma Tor Vergata, Via della Ricerca Scientifica 1, I-00133 Rome, Italy \\
E-mail: \email{martin.hansen@roma2.infn.it}
}

\author{
Kasper Langæble \\
CP3-Origins, University of Southern Denmark, Campusvej 55, DK-5230 Odense M, Denmark \\
E-mail: \email{langaeble@cp3.sdu.dk}
}

\author{
Francesco Sannino \\
CP3-Origins, University of Southern Denmark, Campusvej 55, DK-5230 Odense M, Denmark \\
E-mail: \email{sannino@cp3.sdu.dk}
}

\abstract{
We present an extension of chiral perturbation theory that explicitly includes an isosinglet scalar in the Lagrangian. The dynamical effects from the scalar state is of phenomenological relevance in theories where the mass of the isosinglet scalar is comparable to the mass of the pseudo-Goldstone bosons. This near-degeneracy of states is for example observed in certain near-conformal BSM models. From the Lagrangian we calculate the one-loop radiative corrections to the pion mass, the pion decay constant, and the scalar mass. We then proceed and use the results to fit numerical lattice data for an SU(3) gauge theory with $N_f=8$ light flavours in the fundamental representation.
}

\FullConference{
XIII Quark Confinement and the Hadron Spectrum - Confinement 2018 \\
31 July - 6 August 2018 \\
Maynooth University, Ireland
}

\newcommand{\cO}{\mathcal{O}}
\newcommand{\cL}{\mathcal{L}}
\newcommand{\trace}[1]{\langle{#1}\rangle}
\newcommand{\Jbar}{\overline{J}}
\newcommand{\Hbar}{\overline{H}}

\begin{document}

\begin{figure}[!t]
\begin{center}
\begin{tikzpicture}
 \draw [-{Latex[length=3mm]}] (1,0) -- (1,5); 
 \draw (0.6,0) -- (1.4,0) node[xshift=1mm,anchor=west] {$0$};
 \draw (0.6,0.7) -- (1.4,0.7) node[xshift=1mm,anchor=west] {$\pi$};
 \draw (0.6,2.25) -- (1.4,2.25) node[xshift=1mm,anchor=west] {$\sigma$};
 \draw (0.6,3.85) -- (1.4,3.85) node[xshift=1mm,anchor=west] {$\rho$};
 \draw [-{Latex[length=3mm]}] (6,0) -- (6,5); 
 \draw (5.6,0) -- (6.4,0) node[xshift=1mm,anchor=west] {$0$};
 \draw (5.6,0.7) -- (6.4,0.7) node[xshift=1mm,anchor=west] {$\pi$};
 \draw (5.6,1.05) -- (6.4,1.05) node[xshift=1mm,anchor=west] {$\sigma$};
 \draw (5.6,3.85) -- (6.4,3.85) node[xshift=1mm,anchor=west] {$\rho$};
 \draw [-{Latex[length=3mm]}] (11,0) -- (11,5); 
 \draw (10.6,0) -- (11.4,0) node[xshift=1mm,anchor=west] {$0$};
 \draw (10.6,0.7) -- (11.4,0.7) node[xshift=1mm,anchor=west] {$\pi$};
 \draw (10.6,2.20) -- (11.4,2.20) node[xshift=1mm,anchor=west] {$\sigma$};
 \draw (10.6,2.55) -- (11.4,2.55) node[xshift=1mm,anchor=west] {$K$};
 \draw (10.6,3.85) -- (11.4,3.85) node[xshift=1mm,anchor=west] {$\rho$};
 \node[anchor=center] at (1,-1) {2-flavor QCD};
 \node[anchor=center] at (6,-1) {Near-conformal BSM};
 \node[anchor=center] at (11,-1) {3-flavor QCD};
\end{tikzpicture}
\end{center}
\caption{Hierarchy of masses for different models. In 2-flavour QCD there is a large gap between the pions and the heavier states, while in certain near-conformal BSM models, the pions are observed to be almost degenerate with the isosinglet scalar. Similarly, in the case of 3-flavour QCD, the isosinglet scalar is almost degenerate with the kaons.}
\label{fig:scales}
\end{figure}
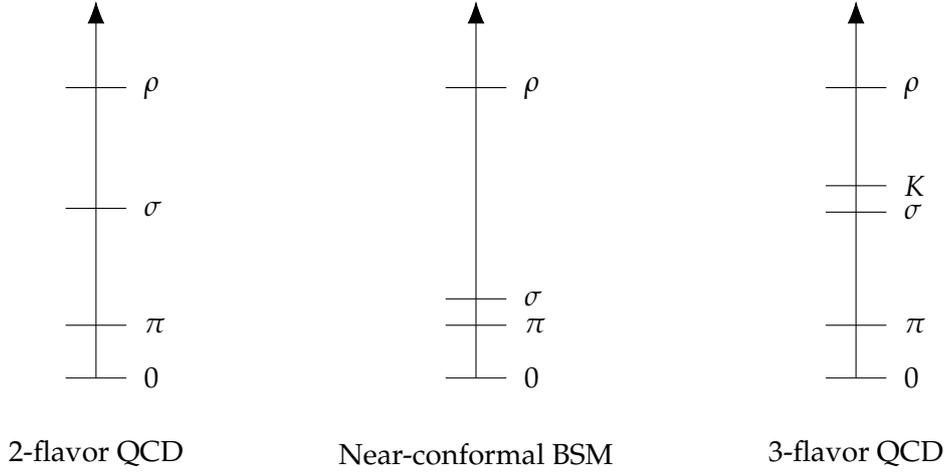

\section{Introduction}
In recent years there has been a renewed interest in studying the properties and the phenomenology of the isosinglet scalar, both in QCD and in near-conformal BSM models, albeit for different reasons. In the former case, new advances in lattice field theory have made it possible to study the isosinglet channel in scattering processes, revealing many interesting and non-trivial properties \cite{Briceno:2016mjc,Briceno:2017qmb}. In the latter case, the isosinglet scalar is interesting because of its relation to conformal symmetry. In particular, if the model of interest is sufficiently close to the conformal window, but still in the chirally broken phase, the isosinglet scalar can (at least partially) be identified with the Goldstone boson arising from the spontaneous breaking of scale invariance. In this case, the scalar is commonly known as a dilaton, and because of its origin as a Goldstone boson, it might potentially be very light. This has indeed been observed in lattice simulations \cite{Aoki:2013zsa,Aoki:2016wnc,Appelquist:2018yqe,Aoki:2017fnr,Fodor:2012ty,Fodor:2015vwa,Fodor:2016pls}, where in some cases the isosinglet scalar is observed to be lighter than the pions. However, it should be mentioned that, assuming these BSM models indeed are in the chirally broken phase, sufficiently close the chiral limit, the pions will always be the lightest states.

At low energy, strongly coupled theories are described by chiral perturbation theory, under the assumption that there is a gap between the mass of the Goldstone bosons and the heavier states, such that these heavy states can be integrated out. As shown on Fig.~\ref{fig:scales}, in two-flavour QCD this is indeed the case, but in the previously mentioned case of near-conformal BSM models, there might not be any separation between the isosinglet scalar and the pions. Furthermore, when considering three-flavour QCD, the kaons are almost degenerate with the isosinglet scalar, which again shows a lack of proper separation.

Due to this lack of separation between the isosinglet scalar and the Goldstone bosons, in some cases chiral perturbation theory might not be a reliable description of the low energy physics. For this reason, there have been several attempts at constructing effective field theories that take both of these states into account \cite{Soto:2011ap,Ametller:2014vba,Hansen:2016fri,Golterman:2016lsd,Golterman:2016cdd,Golterman:2018mfm,Appelquist:2017wcg,Appelquist:2017vyy,Appelquist:2018tyt}. While the different theories incorporate the isosinglet scalar in different ways, they are all extensions of either chiral perturbation theory or the linear sigma model. In these proceedings we will discuss an extension of chiral perturbation theory described in \cite{Hansen:2016fri}.

\section{Approach}
In this section we introduce a version of chiral perturbation theory augmented with an isosinglet scalar field. To keep the discussion short, we will only describe the main points, while the details can be found in the original paper \cite{Hansen:2016fri}.

Before we can write down the Lagrangian, we have to adopt an appropriate counting scheme that includes both the pion and the scalar mass. To this end, we choose the simplest possible extension, where the scalar counts in the same way as the pions
\begin{equation}
 \cO(m_\pi^2) = \cO(m_\sigma^2) = \cO(p^2)~.
\end{equation}
There are two reasons for choosing this particular counting scheme. First of all, when choosing a counting scheme where the scalar counts differently from the pions, in the perturbative expansion, a Feynman diagram including both scalars and pions, does no longer contribute to a single order in the chiral expansion, but in fact to different neighbouring orders. While this is only a mathematical problem, it does make the calculations more complicated. The second reason is phenomenologically motivated, because in lattice simulations of near-conformal BSM models, it is observed that the scalar mass in fact does seem to scale similarly to the pion mass. This is of course only true in some intermediate range of quark masses, because in the chiral limit the pions are massless, while the scalar is not, i.e.
\begin{equation}
 m_\pi^2 = Am_q~,\qquad m_\sigma^2 = m_0^2 + Bm_q~.
\end{equation}
Here $m_0$ is the scalar mass in the chiral limit and $m_q$ is the quark mass. This means that, sufficiently close to the chiral limit, one should indeed integrate out the scalar and use normal chiral perturbation theory.

Having established the counting scheme, we can write down the Lagrangian for our effective theory. Let $G$ be the global flavour symmetry and let $H$ denote the stability group after spontaneous chiral symmetry breaking. The Goldstone boson manifold $G/H$ is then parametrized by
\begin{equation}
 u = \exp\left(\frac{i}{\sqrt{2}f_\pi}\phi^aX^a\right)~,
\end{equation}
where $f_\pi$ is the tree-level pion decay constant and $X^a$ are the broken generators. From this definition we can build the two primary invariants (i.e.~objects invariant under the stability group) used to construct the Lagrangian
\begin{align}
u_\mu &= i(u^\dagger(\partial_\mu-ir_\mu)u-u(\partial_\mu-il_\mu)u^\dagger)~, \\
\chi_\pm &= u^\dagger\chi u^\dagger \pm u\chi^\dagger u~.
\end{align}
The first invariant $u_\mu$ is used to construct the kinetic term, with $l_\mu$ and $r_\mu$ being the external currents. The second invariant $\chi_\pm$ is used to construct the mass term, with $\chi$ the diagonal mass matrix. In fact, at leading order the chiral Lagrangian is simply given by
\begin{equation}
 \cL_2 = \frac{f_\pi^2}{4}\trace{u_\mu u^\mu + \tilde\chi_+}~,
\end{equation}
where $\tilde\chi_+=\chi_+-(\chi+\chi^\dagger)$ is the mass term without the constant piece (this subtraction is needed later on) and $\trace{\cdot}$ denotes the trace in flavour space. At higher order, the chiral Lagrangian contains many more terms, each associated with an unknown low-energy constant (LEC). For example, at next-to-leading order (NLO) we have terms like
\begin{equation}
 \cL_4 = L_0\trace{u_\mu u_\nu u^\mu u^\nu} + L_1\trace{u_\mu u^\mu}\trace{u_\nu u^\nu} + L_2\trace{u_\mu u_\nu}\trace{u^\mu u^\nu} + \cdots~.
\end{equation}
The exact expression for the Lagrangian is not important for the current discussion and we refer to \cite{Bijnens:2009qm,Hansen:2016fri} for details.  While the chiral Lagrangian only depends on the global flavour symmetry, the LECs encode information about the underlying strong dynamics. For this reason, they can be divided into contributions from various sources, such as heavier resonances
\begin{equation}
 L_i = \hat{L}_i + \sum_R L_i^R~.
\end{equation}
Here $L_i^R$ is the contribution from a resonance $R$ and $\hat{L}_i$ is a remainder not directly related to any resonance. For example, under the assumption of vector meson dominance, the contributions $L_i^R$ can be written in terms of the decay constants and masses of these heavy vector resonances \cite{Gasser:1983yg,Ecker:1988te}. In the same way, the isosinglet scalar might contribute to the LECs, but when the scalar is light, the contribution is dynamical and not just a constant.

With the previous discussion in mind, we will now introduce the isosinglet scalar $\sigma$ in the chiral Lagrangian as a non-trivial background field \cite{Soto:2011ap,Cacciapaglia:2014uja}. In practice this is done by expanding each coefficient in the Lagrangian in powers of $\sigma/f_\pi$. Because we are interested in calculating the radiative corrections to the two-point functions at next-to-leading order, the expansion is only needed for the leading-order Lagrangian and we can stop the series expansion at second order.
\begin{align}
\begin{split}
 \cL_2
 &~=~ \frac{f_\pi^2}{4}\left[1+S_1\left(\frac{\sigma}{f_\pi}\right)+S_2\left(\frac{\sigma}{f_\pi}\right)^2+\cdots\right]\trace{u_\mu u^\mu}  \\
 &~+~ \frac{f_\pi^2}{4}\left[1+S_3\left(\frac{\sigma}{f_\pi}\right)+S_4\left(\frac{\sigma}{f_\pi}\right)^2+\cdots\right]\trace{\tilde{\chi}_+}
\end{split}
\end{align}
It is now evident that the subtraction in the mass term is needed to avoid terms that only include the scalar field. The associated Lagrangian for the scalar field can be written as
\begin{equation}
 \cL_\sigma
 = \frac{1}{2}\partial_\mu\sigma\partial^\mu\sigma
 - \frac{1}{2}m_\sigma^2\sigma^2\left[1+S_5\left(\frac{\sigma}{f_\pi}\right)+S_6\left(\frac{\sigma}{f_\pi}\right)^2\right].
\end{equation}
In principle we should also perform a series expansion in front of the kinetic term, but since we will only consider on-shell quantities, these terms are related to the potential via the equations of motion. Because the scalar field parametrise the fluctuations around the vacuum, it must have vanishing expectation value, and this leads to certain constraints on the two parameters $S_5$ and $S_6$ controlling the potential
\begin{equation}
 S_5 \geq-2\sqrt{S_6}~,\qquad S_6\geq0~.
 \label{eq:s5s6}
\end{equation}
On top of these definitions, we also need a set of counterterms to renormalize the Lagrangian. These can be found in the original paper \cite{Hansen:2016fri} together with a detailed description of the renormalization procedure.

Before continuing, we want to mention that our approach is completely generic, within the limits of the chosen counting scheme. We make no assumption about the nature of the scalar, but one can show that different physical origins correspond to imposing constraints on the couplings $S_i$ between the scalar and the pions, as shown later on. Moreover, when performing the calculations at NLO, the results are largely independent of the pattern of chiral symmetry breaking, and for this reason, the results can easily be applied to any model of interest.

\section{Results\label{sec:results}}
\begin{figure}[t]
 \centering
 \includegraphics[scale=0.8]{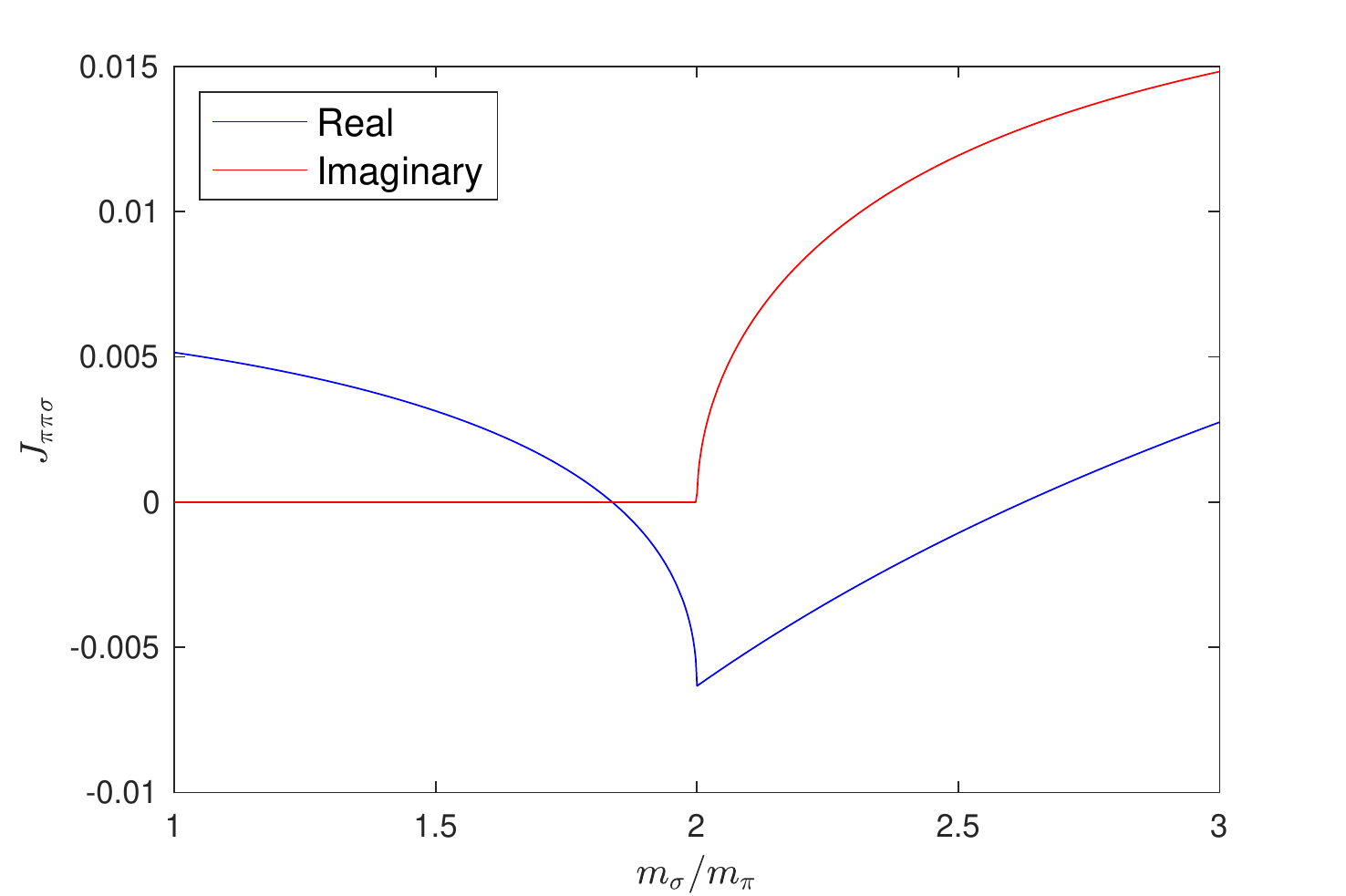}
 \caption{Behaviour of $J_{\pi\pi\sigma}$ as a function of the ratio $m_\sigma/m_\pi$. A branch cut is developed when $m_\sigma=2m_\pi$ after which the imaginary part describes the decay width of the isosinglet scalar in the $\sigma\to\pi\pi$ channel.}
 \label{fig:cut}
\end{figure}

Having defined the Lagrangian, we are now able to calculate the two-point functions needed to define the renormalized pion mass, pion decay constant, and scalar mass, at next-to-leading order. For both the pion mass and the pion decay constant there are four diagrams in total; two diagrams with scalars in the loop, one contact term and one diagram only with pions. We define the renormalized pion mass as the pole mass in the propagator, and the result reads
\begin{align}
\begin{split}
 \hat{m}_\pi^2
 = m_\pi^2
 + \frac{m_\pi^4}{f_\pi^2}(a_1+a_2L_\pi+a_3J_{\pi\sigma\pi})
 &+ \frac{m_\sigma^4}{f_\pi^2}(a_4L_\sigma + a_5J_{\pi\sigma\pi}) \\
 &\quad+ \frac{m_\pi^2m_\sigma^2}{f_\pi^2}(a_6 + a_7L_\pi+a_8L_\sigma+a_9J_{\pi\sigma\pi})~,
\end{split}
\end{align}
while the result for the pion decay constant reads
\begin{align}
\begin{split}
 \hat{f}_\pi
 = f_\pi + \frac{m_\pi^2}{f_\pi}(b_1 + b_2L_\pi + b_3J_{\pi\sigma\pi})
 &+ \frac{m_\sigma^2}{f_\pi}(b_4 + b_5L_\sigma + b_6J_{\pi\sigma\pi}) \\
 &\quad+ \frac{H_{\pi\sigma\pi}}{f_\pi}(b_7m_\pi^4 + b_8m_\sigma^4 + b_9m_\pi^2m_\sigma^2)~.
 \label{eq:Fhat}
\end{split}
\end{align}
Here $a_i$ and $b_i$ are specific combinations of the various low-energy constants. In the case of normal chiral perturbation theory, the only non-zero constants are $a_{1,2}$ and $b_{1,2}$ which means that including the scalar vastly increases the complexity of the results. In the equations we use the following auxiliary functions as shorthand notation for the chiral logs and the unitarity corrections. We refer to the appendix of \cite{Hansen:2016fri} for the definition of the barred functions.
\begin{align}
\begin{split}
  L_x &= \frac{1}{16\pi^2}\log\left(\frac{m_x^2}{\mu^2}\right) \\
 J_{xyz} &= \frac{1}{16\pi^2}\Bigg[\Jbar(m_x^2,m_y^2,m_z^2)+1\Bigg] \\
 H_{xyz} &= \frac{1}{16\pi^2}\Bigg[\Hbar(m_x^2,m_y^2,m_z^2)\Bigg]
 \label{eq:LJH}
\end{split}
\end{align}
We also calculated the renormalized scalar mass, which reads
\begin{align}
\begin{split}
 \hat{m}_\sigma^2
 = m_\sigma^2
 + \frac{m_\sigma^4}{f_\pi^2}(c_1L_\sigma + c_2J_{\pi\pi\sigma} + c_3J_{\sigma\sigma\sigma})
 &+ \frac{m_\pi^4}{f_\pi^2}(c_4L_\pi + c_5J_{\pi\pi\sigma}) \\
 &\quad+ \frac{m_\pi^2m_\sigma^2}{f_\pi^2}(c_6L_\pi + c_7J_{\pi\pi\sigma})~.
\end{split}
\end{align}
The calculation of the scalar self-energy includes a diagram with an intermediate pion loop. Because of this diagram, when the scalar is sufficiently heavy, the pions are able to go on-shell, corresponding to the $\sigma\to\pi\pi$ decay channel being kinematically allowed. In the analytical expression for the renormalized scalar mass, this results in a branch cut in the function $J_{\pi\pi\sigma}$ as shown in Fig.~\ref{fig:cut}. The branch cut starts at $m_\sigma=2m_\pi$ and above this threshold the decay width of the scalar can be extracted from the imaginary part
\begin{equation}
 \Gamma = \frac{n_\pi}{16\pi m_\sigma f_\pi^2}\left(S_1\left(\frac{m_\sigma^2}{2}-m_\pi^2\right)+S_3m_\pi^2\right)^2\sqrt{1-\frac{4m_\pi^2}{m_\sigma^2}}~.
\end{equation}
Here $n_\pi$ is the number of pions for the given pattern of chiral symmetry breaking and $S_1$ and $S_3$ are the two low-energy constants parametrizing the decay width. Because $S_3$ parametrize the interaction between the scalar and the pion mass term, in the chiral limit, this coefficient is irrelevant, and the decay width only depends on $S_1$.

\subsection{Consistency checks}
We performed several consistency checks of the previous results to ensure their validity. For the pion we checked that the renormalized mass $\hat{m}_\pi^2$ vanishes in the limit chiral limit where $m_\pi^2\to0$. This is a non-trivial check, because the $m_\sigma^4$ term only vanishes due to an exact cancellation in the chiral limit.

For the pion decay constant we checked that we obtain a finite and non-zero value in the chiral limit, which indeed is the case
\begin{equation}
 \hat{f}_\pi = f_\pi + \frac{m_\sigma^2}{f_\pi}\left(b_4 + (b_5+b_6)L_\sigma - \frac{b_8}{32\pi^2}\right).
\end{equation}
Because of the scalar corrections, we observe that $\hat{f}_\pi$ and $f_\pi$ no longer coincide in the chiral limit. In fact, the entire right-hand side corresponds to what is denoted $f_\pi$ in normal chiral perturbation theory.

Finally we checked that all the results are independent of the renormalization scale. This means that changing the renormalization scale corresponds to a shift in all the LECs. Again, this is a non-trivial property, because it depends on the specific combinations of the functions defined in Eq.~\eqref{eq:LJH}.

\section{Origins of the scalar}
As already mentioned, our approach for including the scalar is completely generic. However, if we assume a specific physical origin for the scalar, we can make predictions for the couplings $S_i$ between the scalar and the pions. As an example, here we will consider the case where the scalar emerges as a pseudo-dilaton \cite{Goldberger:2008zz,Matsuzaki:2013eva}. In this scenario, the scalar is introduced as the conformal compensator, and the Lagrangian reads
\begin{equation} 
 \cL_2 = \frac{f_\pi^2}{4}\left[\trace{u_\mu u^\mu}\exp\left(\frac{2\sigma}{f_\pi}\right) + \trace{\chi_+}\exp\left(\frac{y\sigma}{f_\pi}\right)\right].
\end{equation}
Although we use $f_\pi$ as the compensating scale for the pseudo-dilaton in the exponential, de facto, depending on the microscopic realization it can differ, but our results still apply. Expanding the exponential to second order we find that our couplings are given by
\begin{equation}
 S_1 = S_2 = 2~,\qquad S_3=y~,\qquad S_4 = \frac{y^2}{2}~.
\end{equation}
Here $y=3-\gamma^*$ with $\gamma^*$ being the anomalous dimension of the fermion mass in the underlying gauge theory. It is now evident that $\gamma^*$ is the only new parameter in the expression for the pion mass and the pion decay constant, when the scalar field is a pseudo-dilaton.

With this example it is evident that the value of the couplings $S_i$ can be used to make predictions about the origin of the scalar field. The original paper \cite{Hansen:2016fri} contains a few more examples of different physical origins.

\section{Fitting lattice data}
\begin{table}
\centering
\def\arraystretch{1.1}
\begin{tabular}{c|ccc|cccc}
 $m_q$ & $L^3\times T$& $\hat{m}_\sigma$ & Table & $L^3\times T$& $\hat{m}_\pi$ & $\hat{f}_\pi$ & Table \\
 \hline\hline
  0.012 & $42^3\times56$ & 0.151(27) & XVII & $42^3\times56$ & 0.16362(43) & 0.04542(27) & XXI   \\
  0.015 & $36^3\times48$ & 0.162(59) & XVII & $42^3\times56$ & 0.18614(44) & 0.05054(15) & XXI   \\
  0.020 & $36^3\times48$ & 0.190(36) & XVII & $36^3\times48$ & 0.22052(33) & 0.05848(15) & XXII  \\
  0.030 & $30^3\times40$ & 0.282(39) & XVII & $36^3\times48$ & 0.28084(39) & 0.07137(20) & XXII  \\
  0.040 & $30^3\times40$ & 0.365(51) & XVII & $30^3\times40$ & 0.33501(21) & 0.08264(10) & XXIII \\
  0.060 & $24^3\times32$ & 0.46(13)  & XVII & $30^3\times40$ & 0.43035(44) & 0.10118(28) & XXIII \\
\end{tabular}
\caption{Numerical data used for the fitting procedure. The table column refer to the table in \cite{Aoki:2016wnc} from where the data was taken. For simplicity we averaged the upper and lower errors for the scalar mass.}
\label{tab:data}
\end{table}

As an example, we will now use the results from section \ref{sec:results} to fit a set of lattice data from the LatKMI collaboration. The simulated model is an SU(3) gauge theory with $N_f=8$ dynamical flavours \cite{Aoki:2016wnc}. This model is an example of a near-conformal BSM model, and (although not entirely conclusive) it is believed to be in the chirally broken phase. Furthermore, for this model the pion and the scalar mass are almost degenerate over the explored range of quark masses, which was the assumption used for choosing our counting scheme.

The numerical data used for the fit is shown in Table~\ref{tab:data}, together with references to where the data was found in \cite{Aoki:2016wnc}. Because our extension allows us to simultaneously fit the pion mass, the pion decay constant, and the scalar mass, ideally all of these quantities should have similar uncertainties to properly constrain the fit. However, due to the difficulty in measuring the scalar state in lattice simulations, this quantity will always have a significantly larger uncertainty.

\begin{figure}[t]
 \centering
 \includegraphics[scale=0.9]{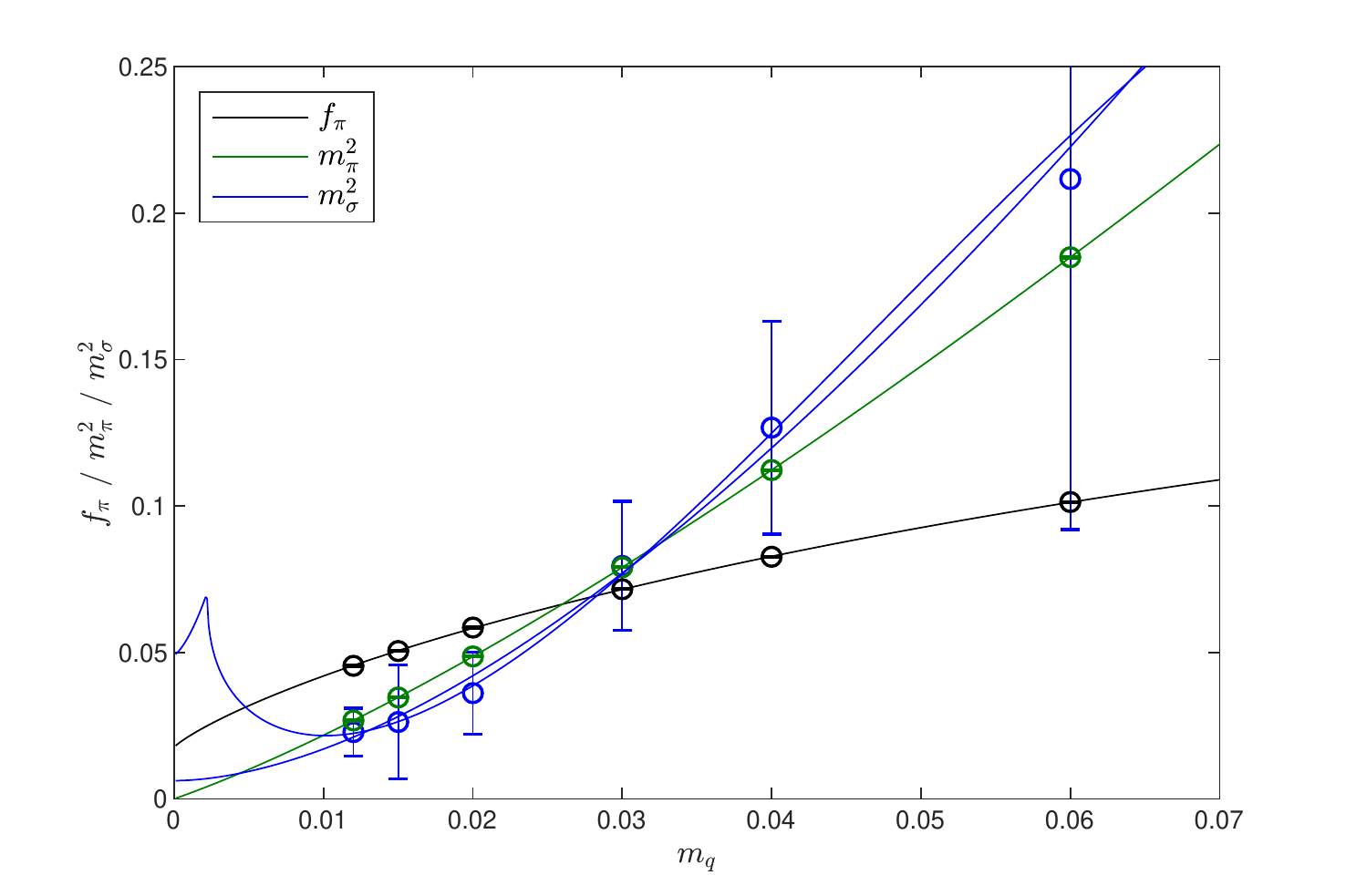}
 \caption{Fit to the numerical data in Table~\ref{tab:data}. Two different minima were found with the same value of $\chi^2$. For the two minima there is no visual difference for the pion quantities, but there is a small difference for the scalar mass, because this quantity has significantly larger errorbars.}
 \label{fig:fit}
\end{figure}

The list in Eq.~\eqref{eq:list} contains the 13 free parameters used for the fit. In the continuum the two coefficients $a_M$ and $a_F$ are known numbers, but at non-zero lattice spacing these will receive unknown corrections. For this reason, it is usually very difficult to fit lattice data when using the continuum values, and this is still true even with the additional parameters introduced by the scalar extension. This is why we also include these as free parameters.
\begin{equation}
 \{B_0~,~f_\pi~,~a_M~,~b_M~,~a_F~,~b_F~,~m_\sigma~,~S_1~,~S_2~,~S_3~,~S_4~,~S_5~,~S_6\}
 \label{eq:list}
\end{equation}
The result of the fit is shown in Fig.~\ref{fig:fit}. During the fitting procedure we found two different minima with the same value of $\chi^2/\mathrm{dof}=1.39$. For both of these minima, the visual result is the same for the pion mass and the pion decay constant, while the visual result for the scalar is slightly different. This is easily understood, because the uncertainty on the scalar mass is significantly larger, such that the value of $\chi^2$ is completely determined by the pion quantities. Close to the chiral limit, the fit for the scalar mass is very different for the two minima, however, this is not important because the scalar is unstable in this region. As a consequence, the fit cannot be used to extrapolate the scalar mass, and we are satisfied that the fit for the scalar mass is consistent in the region of quark masses where we have data.

For one of the minima, we see the branch cut in the scalar mass close to the chiral limit, while in the other case, the coefficients in front of the $J_{\pi\pi\sigma}$ function are so small that we do not see the branch cut. The fact that the coefficients are quite different for the two minima proves that the uncertainty on the scalar mass is too large to properly constrain the $S_i$ parameters. This is an important observation because, as previously discussed, the values of the fitted parameters can be used to distinguish between different physical origins of the scalar, but unfortunately this is not possible with the currently available data.

We finally remark that, for both minima, the constraint on the scalar potential in Eq. \eqref{eq:s5s6} is satisfied because both coefficients are positive. Furthermore, in the chiral limit the difference between the renormalized and the bare pion decay constant is relatively small, namely:
\begin{equation}
 \frac{\hat{f}_\pi-f_\pi}{\hat{f}_\pi} \sim 5\%
\end{equation}
This is expected when the scalar only acts as a small perturbation of chiral perturbation theory.

\section{Conclusion}
We presented a simple extension of chiral perturbation theory that accounts for the dynamical effects of a light isosinglet scalar state. After discussing the chosen counting scheme, we introduce the Lagrangian and calculate the radiative one-loop corrections to the pion mass, the pion decay constant, and the scalar mass. Our approach is very generic, it makes no assumptions about the physical origin of the scalar and the results are valid for different patterns of chiral symmetry breaking. For this reason, the framework can be used for a large class of interesting models. After presenting the results, we argue that different physical origins of the scalar correspond to imposing constraints on some of the low-energy constants, and as such, in principle one can make predictions about the nature of the scalar by fitting these constants to data. For this reason, we use the results to fit numerical data from a lattice simulation, and while this is possible, the uncertainty on the scalar mass is too large to properly constrain the interesting coefficients.


\begin{thebibliography}{99}

\bibitem{Briceno:2016mjc} 
  R.~A.~Briceno, J.~J.~Dudek, R.~G.~Edwards and D.~J.~Wilson,
 \emph{Isoscalar $\pi\pi$ scattering and the $\sigma$ meson resonance from QCD},
  Phys.\ Rev.\ Lett.\  {\bf 118}, no. 2, 022002 (2017),
  arXiv:1607.05900 (hep-ph)



\bibitem{Briceno:2017qmb} 
  R.~A.~Briceno, J.~J.~Dudek, R.~G.~Edwards and D.~J.~Wilson,
 \emph{Isoscalar $\pi\pi, K\overline{K}, \eta\eta$ scattering and the $\sigma, f_0, f_2$ mesons from QCD},
  Phys.\ Rev.\ D {\bf 97}, no. 5, 054513 (2018),
  arXiv:1708.06667 (hep-lat)



\bibitem{Aoki:2013zsa} 
  Y.~Aoki {\it et al.} [LatKMI Collaboration],
  \emph{Light composite scalar in twelve-flavor QCD on the lattice},
  Phys.\ Rev.\ Lett.\  {\bf 111}, no. 16, 162001 (2013),
  arXiv:1305.6006 (hep-lat)



\bibitem{Aoki:2016wnc} 
  Y.~Aoki {\it et al.} [LatKMI Collaboration],
  \emph{Light flavor-singlet scalars and walking signals in $N_f=8$ QCD on the lattice},
  Phys.\ Rev.\ D {\bf 96}, no. 1, 014508 (2017),
  arXiv:1610.07011 (hep-lat)



\bibitem{Appelquist:2018yqe} 
  T.~Appelquist {\it et al.} [Lattice Strong Dynamics Collaboration],
  \emph{Nonperturbative investigations of SU(3) gauge theory with eight dynamical flavors},
  arXiv:1807.08411 (hep-lat)



\bibitem{Aoki:2017fnr} 
  Y.~Aoki {\it et al.},
 \emph{Flavor-singlet spectrum in multi-flavor QCD},
  EPJ Web Conf.\  {\bf 175}, 08023 (2018),
  arXiv:1710.06549 (hep-lat)



\bibitem{Fodor:2012ty} 
  Z.~Fodor, K.~Holland, J.~Kuti, D.~Nogradi, C.~Schroeder and C.~H.~Wong,
  \emph{Can the nearly conformal sextet gauge model hide the Higgs impostor?},
  Phys.\ Lett.\ B {\bf 718}, 657 (2012),
  arXiv:1209.0391 (hep-lat)



\bibitem{Fodor:2015vwa} 
  Z.~Fodor, K.~Holland, J.~Kuti, S.~Mondal, D.~Nogradi and C.~H.~Wong,
  \emph{Toward the minimal realization of a light composite Higgs},
  PoS LATTICE {\bf 2014}, 244 (2015),
  arXiv:1502.00028 (hep-lat)



\bibitem{Fodor:2016pls} 
  Z.~Fodor, K.~Holland, J.~Kuti, S.~Mondal, D.~Nogradi and C.~H.~Wong,
  \emph{Status of a minimal composite Higgs theory},
  PoS LATTICE {\bf 2015}, 219 (2016),
  arXiv:1605.08750 (hep-lat)



\bibitem{Soto:2011ap} 
  J.~Soto, P.~Talavera and J.~Tarrus,
  \emph{Chiral Effective Theory with A Light Scalar and Lattice QCD},
  Nucl.\ Phys.\ B {\bf 866}, 270 (2013),
  arXiv:1110.6156 (hep-ph)



\bibitem{Ametller:2014vba} 
  L.~Ametller and P.~Talavera,
  \emph{Lowest resonance in QCD from low-energy data},
  Phys.\ Rev.\ D {\bf 89}, no. 9, 096004 (2014),
  arXiv:1402.2649 (hep-ph)



\bibitem{Hansen:2016fri} 
  M.~Hansen, K.~Langæble and F.~Sannino,
  \emph{Extending Chiral Perturbation Theory with an Isosinglet Scalar},
  Phys.\ Rev.\ D {\bf 95}, no. 3, 036005 (2017),
  arXiv:1610.02904 (hep-ph)



\bibitem{Golterman:2016lsd} 
  M.~Golterman and Y.~Shamir,
  \emph{Low-energy effective action for pions and a dilatonic meson},
  Phys.\ Rev.\ D {\bf 94}, no. 5, 054502 (2016),
  arXiv:1603.04575 (hep-ph)



\bibitem{Golterman:2016cdd} 
  M.~Golterman and Y.~Shamir,
  \emph{Effective pion mass term and the trace anomaly},
  Phys.\ Rev.\ D {\bf 95}, no. 1, 016003 (2017),
  arXiv:1611.04275 (hep-ph)



\bibitem{Golterman:2018mfm} 
  M.~Golterman and Y.~Shamir,
  \emph{The large-mass regime of the dilaton-pion low-energy effective theory},
  Phys.\ Rev.\ D {\bf 98}, 056025 (2018),
  arXiv:1805.00198 (hep-ph)



\bibitem{Appelquist:2017wcg} 
  T.~Appelquist, J.~Ingoldby and M.~Piai,
  \emph{Dilaton EFT Framework For Lattice Data},
  JHEP {\bf 1707}, 035 (2017),
  arXiv:1702.04410 (hep-ph)



\bibitem{Appelquist:2017vyy} 
  T.~Appelquist, J.~Ingoldby and M.~Piai,
  \emph{Analysis of a Dilaton EFT for Lattice Data},
  JHEP {\bf 1803}, 039 (2018),
  arXiv:1711.00067 (hep-ph)



\bibitem{Appelquist:2018tyt} 
  T.~Appelquist {\it et al.} [LSD Collaboration],
  \emph{Linear Sigma EFT for Nearly Conformal Gauge Theories},
  arXiv:1809.02624 (hep-ph)



\bibitem{Bijnens:2009qm} 
  J.~Bijnens and J.~Lu,
  \emph{Technicolor and other QCD-like theories at next-to-next-to-leading order},
  JHEP {\bf 0911}, 116 (2009),
  arXiv:0910.5424 (hep-ph)



\bibitem{Gasser:1983yg} 
  J.~Gasser and H.~Leutwyler,
  \emph{Chiral Perturbation Theory to One Loop},
  Annals Phys.\  {\bf 158}, 142 (1984)



\bibitem{Ecker:1988te} 
  G.~Ecker, J.~Gasser, A.~Pich and E.~de Rafael,
  \emph{The Role of Resonances in Chiral Perturbation Theory},
  Nucl.\ Phys.\ B {\bf 321}, 311 (1989)



\bibitem{Cacciapaglia:2014uja} 
  G.~Cacciapaglia and F.~Sannino,
  \emph{Fundamental Composite (Goldstone) Higgs Dynamics},
  JHEP {\bf 1404}, 111 (2014),
  arXiv:1402.0233 (hep-ph)



\bibitem{Goldberger:2008zz} 
  W.~D.~Goldberger, B.~Grinstein and W.~Skiba,
  \emph{Distinguishing the Higgs boson from the dilaton at the Large Hadron Collider},
  Phys.\ Rev.\ Lett.\  {\bf 100}, 111802 (2008),
  arXiv:0708.1463 (hep-ph)



\bibitem{Matsuzaki:2013eva} 
  S.~Matsuzaki and K.~Yamawaki,
  \emph{Dilaton Chiral Perturbation Theory: Determining the Mass and Decay Constant of the Technidilaton on the Lattice},
  Phys.\ Rev.\ Lett.\  {\bf 113}, no. 8, 082002 (2014),
  arXiv:1311.3784 (hep-lat)





\end{thebibliography}
\end{document}